\documentclass[twocolumn,showpacs,superscriptaddress,amssymb,pre]{revtex4}

\usepackage{graphicx}
\usepackage{color}
\usepackage{dcolumn}
\usepackage{bm}
\usepackage{amsmath}
\usepackage[normalem]{ulem}
\usepackage[]{epsfig}
\newcommand{\tw}{t_\mathrm{w}}

\newcommand{\phic}{\varphi_\mathrm{c}}
\renewcommand{\phi}{\varphi}
\newcommand{\ta} {\tau_{\alpha}}
\newcommand{\taua}{\tau_{\alpha}}
\newcommand{\be}{\begin{equation}}
\newcommand{\ee}{\end{equation}}

\begin{document}

\title{Dynamic light scattering measurements in
the activated regime of
dense colloidal hard spheres}

\author{D. El Masri}
\affiliation{Laboratoire des Collo{\"\i}des, Verres et
Nanomat{\'e}riaux, UMR 5587, Universit{\'e} Montpellier II and CNRS,
34095 Montpellier, France}

\author{G. Brambilla}
\affiliation{Laboratoire des Collo{\"\i}des, Verres et
Nanomat{\'e}riaux, UMR 5587, Universit{\'e} Montpellier II and CNRS,
34095 Montpellier, France}

\author{M. Pierno}
\affiliation{Laboratoire des Collo{\"\i}des, Verres et
Nanomat{\'e}riaux, UMR 5587, Universit{\'e} Montpellier II and CNRS,
34095 Montpellier, France}

\author{G. Petekidis}
\affiliation{IESL-FORTH and Department of Material Science and
Technology, University of Crete, GR-711 10 Heraklion, Greece}

\author{A. B. Schofield}
\affiliation{School of Physics, Edinburgh University, Mayfield Road,
Edinburgh, EH9 3JZ, United Kingdom}

\author{L. Berthier}
\affiliation{Laboratoire des Collo{\"\i}des, Verres et
Nanomat{\'e}riaux, UMR 5587, Universit{\'e} Montpellier II and CNRS,
34095 Montpellier, France}

\author{L. Cipelletti}
\affiliation{Laboratoire des Collo{\"\i}des, Verres et
Nanomat{\'e}riaux, UMR 5587, Universit{\'e} Montpellier II and CNRS,
34095 Montpellier, France}

\date{\today}
\begin{abstract}{We use dynamic light scattering and numerical simulations
to study the approach to equilibrium and the equilibrium dynamics of
systems of colloidal hard spheres over a broad range of density,
from dilute systems up to very concentrated suspensions undergoing
glassy dynamics. We discuss several experimental issues
(sedimentation, thermal control, non-equilibrium aging effects,
dynamic heterogeneity) arising when very large relaxation times are
measured. When analyzed over more than seven decades in time, we
find that the equilibrium relaxation time, $\ta$, of our system is
described by the algebraic divergence predicted by mode-coupling
theory over a window of about three decades. At higher density,
$\ta$ increases exponentially with distance to a critical volume
fraction $\phi_0$ which is much larger than the mode-coupling
singularity. This is reminiscent of the behavior of molecular
glass-formers in the activated regime. We compare these results to
previous work, carefully discussing crystallization and size
polydispersity effects. Our results suggest the absence of a genuine
algebraic divergence of $\ta$ in colloidal hard spheres.}

\end{abstract}



\maketitle

\section{Introduction}
\label{SEC:intro}

Colloidal particles are increasingly popular as model systems for
investigating the behavior of atomic and molecular
materials~\cite{thebook}. As the typical size of a colloidal
particle is comparable to the wavelength of the visible light, one
can use relatively simple, yet powerful, techniques, such as optical
and confocal microscopy~\cite{HabdasCOCIS2002} or static and dynamic
light scattering~\cite{Berne1976}, to fully characterize their
structural and dynamical properties. In contrast to molecular
systems, the interaction between colloids can easily be tuned from
repulsive (e.g. due to electrostatic interactions), to hard
sphere-like~\cite{PuseyNature1986}, moderately attractive (e.g. due
to depletion forces~\cite{PoonJPCM2002}), or even strongly
attractive (e.g. due to van der Waals attractions). Furthermore,
colloidal particles can now be synthesized in a variety of
non-spherical shapes, and their surface can have non-uniform
physical or chemical properties, opening the way to anisotropic
systems with directional interactions that can be precisely
tailored~\cite{GlotzerNatMat2007}.

This variety of morphologies and surface properties, and the
possibility to study colloidal systems with relative ease, comes
with a price. Unlike atoms, colloids have to be synthesized and the
outcome is usually a polydisperse assembly of particles with
different sizes. Depending
on the phenomenon of interest, polydispersity might be an important
control parameter, e.g. when the precise location of phase
transitions must be determined. Moreover, the very same feature that
makes accessible their typical time and spatial scales---their
comparatively large size---also makes colloidal particles prone to
sedimentation. The density of a colloid typically differs from that
of the solvent in which it is suspended. Therefore, colloids
experience a buoyant force, proportional to $\Delta \rho$, the
difference between the density of the particle and that of the
solvent. Sedimentation (or creaming, if $\Delta \rho<0$) is usually
not a severe issue for isolated, submicron colloids. The situation
is however radically different for colloids that form solid
structures, such as aggregates, crystallites, or glasses, because
the gravitational stress due to a large number of particles may add
up. These effects can be mitigated only partially by matching the
solvent density to that of the
particles~\cite{KegelLangmuir2000,SimeonovaPRL2004}, since a perfect
matching can not be achieved in practice. Finally, a large particle
size implies that microscopic motion occurs on a timescale typically
much larger than for molecular systems. Although clearly an
advantage when single particle motion is investigated via direct
visualization techniques, this is not necessarily so when slow,
collective relaxation phenomena are studied, which can occur on
timescales of several days.

In this paper, we discuss some of these issues in relation with our
investigations by dynamic light scattering (DLS) of dense
suspensions of colloidal hard spheres approaching the colloidal
glass transition~\cite{brambilla}. For hard spheres at thermal
equilibrium, several distinct glass transition scenarios have been
described. In a first line of research, the viscosity or,
equivalently, the timescale for structural relaxation,
$\tau_\alpha(\phi)$, is believed to diverge algebraically: \be
\tau_\alpha (\phi) \sim (\phi_{\rm c}- \phi)^{-\gamma}. \label{mct}
\ee This result is both predicted~\cite{mct} by mode coupling theory
(MCT), and supported by light scattering data~\cite{vanmegen}.
Packing fractions $\phi_c \approx 0.57-0.59$ are the most often
quoted values for the location of this colloidal glass transition. A
truly non-ergodic state is also often reported for larger
$\phi$~\cite{pusey,pusey2,vanmegen,weitz}.

Several alternative scenarios~\cite{free,ken,silvio,zamponi} suggest a
stronger divergence:
\be
\tau_\alpha (\phi) = \tau_{\infty}  \exp
\left[ \frac{C}{(\phi_{0} - \phi)^{\delta} } \right].
\label{vft}
\ee
Equation (\ref{vft}) with $\delta=1$ is frequently used to
account for viscosity data~\cite{chaikin,szamel} because it resembles the
Vogel-Fulcher-Tammann (VFT) form used to fit the viscosity of
molecular glass-formers~\cite{reviewnature}, with temperature
replaced by $\phi$. Moreover, it is theoretically expected on the
basis of free volume arguments~\cite{free}, which lead to the
identification $\phi_0 \equiv \phi_{\rm rcp}$, the random close
packing fraction where osmotic pressure diverges. Kinetic arrest
must occur at $\phi_{\rm rcp}$  (possibly with $\delta = 2$~\cite{ken}),
because all particles block each other at that
density~\cite{pointJ,ken,torquato2}. This is analogous to a
$T=0$ glass transition for molecular systems~\cite{longtom}.

Entropy-based theories and
replica calculations~\cite{silvio,zamponi} predict instead a divergence of
$\ta$ at an ideal glass transition at $\phi_0 < \phi_{\rm rcp}$,
where the configurational entropy vanishes but the pressure is still
finite. This is analogous to a finite temperature ideal
glass transition in molecular glass-formers~\cite{longtom}.
In this context, the connection to dynamical properties is made through
nucleation arguments~\cite{wolynes} yielding Eq.~(\ref{vft}), with
$\delta$ not necessarily equal to unity~\cite{BB}. Here, the relaxation
time does not diverge because particles are in close contact,
but because the configurational entropy counting the number of
metastable states vanishes.

In molecular glass-formers where dynamical slowing down can be
followed over as many as 15 decades, the transition from an MCT
regime, Eq.~(\ref{mct}), to an activated one, Eq.~(\ref{vft}), has
been experimentally demonstrated~\cite{reviewnature}. For colloidal
hard spheres, the situation remains controversial, because dynamic
data are available over a much smaller
range~\cite{pusey,vanmegen,chaikin,Chaikin2}, typically five decades
or less. Crucially, only equilibrium
measurements for $\phi <
\phi_\mathrm{c}$ were reported,
leaving unknown the precise nature and location of
the divergence. A confusing feature of the colloidal glass transition,
therefore, is that the `experimental glass transition', which
we denote $\phi_g$, and which occurs
when equilibrium relaxation times become too large to be confidently measured
experimentally, occurs near the fitted MCT singularity, $\phi_g \approx
\phi_c$, while the two crossovers are well-separated in
molecular systems.

At the theoretical level, there is also an on-going debate about the
nature of the glass transition in colloidal hard spheres systems.
theoretical claims exist that the cutoff mechanism suppressing the
MCT divergence in molecular systems is inefficient in colloids due
to the Brownian nature of the microscopic dynamics, suggesting that
MCT could be virtually exact for colloids~\cite{brownian}. This
viewpoint is challenged by more recent MCT
calculations~\cite{ABL1,ABL2}, and by computer studies of simple
model systems where the MCT transition is similarly avoided both for
stochastic and Newtonian
dynamics~\cite{stochastic1,stochastic2,stochastic3}, directly
emphasizing that the cutoff mechanism of the MCT transition in
molecular systems is different from the one described in
Ref.~\cite{brownian}, and is likely very similar for hard sphere
colloids and molecules. The only physical ingredient missing in
these theoretical works is the inclusion of hydrodynamic
interactions, which are always supposed to play little role at very
large volume fractions. It would be very surprising if hydrodynamic
interactions in colloids could suppress activated processes and make
the MCT predictions exact.

In this article we discuss in detail all these issues. In a short
version of this work~\cite{brambilla}, we claimed that we had been
able to detect ergodic dynamics for a colloidal hard sphere system
at volume fractions above the mode-coupling transition, $\phi_c$,
because our experiment was able to cover an unprecedented range of
variation of the equilibrium relaxation time. In the present
article, we discuss in detail the several challenges we had to face
in order to obtain these data, and we argue that our experimental
results should apply quite generally to colloidal hard spheres. We
first describe our sample and experimental set-up in
Sec.~\ref{SEC:M&M}. In Sec.~\ref{SEC:challenges} we specifically
discuss the difficulties associated with measurements of long
relaxation timescales. In Sec.~\ref{SEC:ANALYSIS} we analyze our
results for the equilibrium dynamics, and compare them with previous
experiments and a new set of numerical simulations. Finally, we
conclude the paper in Sec.~\ref{SEC:Conclusions}.

\section{Colloidal particles and experimental setup}
\label{SEC:M&M}

\subsection{Sample preparation}

The particles are poly(methylmethacrylate) (PMMA) spheres of average
radius $R=130$ nm, sterically stabilized by a thin layer of
poly(12-hydroxystearic) acid of thickness $\approx 10$ nm. They are
physically similar to those used in previous studies of the glass transition
(see e.g.~\cite{PuseyNature1986,vanmegen,chaikin}).
A major advantage is that they are slightly smaller
than in previous studies. This implies that relaxation in the dilute
limit is faster, and that we can cover a broader range of volume fractions.
The particles are polydisperse (standard
deviation of the size distribution normalized by the average size
$\sigma \approx 10 \%$, as obtained by a cumulant analysis of DLS
data~\cite{PuseyJPhysChem1982}), in order to prevent crystallization
for at least several months, much longer than the typical experiment
duration. They are suspended in a mixture of cis/trans-decalin and
tetralin (66/34 w/w), whose refractive index $n_D = 1.503$ at $T =
27~ ^{\circ}\mathrm{C}$ closely matches that of the colloids,
thereby minimizing van der Waals attractions and allowing light
scattering experiments to be performed in the single scattering
regime. Prior to each measurement, the suspensions are vortexed for
about 6 h, centrifuged for a few minutes to remove air bubbles and
then kept on a gently tumbling wheel for at least 12 h. The waiting
time or sample age, $\tw$, is measured from the end of the tumbling.
When several measurements are repeated on a sample at a given volume
fraction, the dynamics are re-initialized by tumbling it on the
wheel overnight, with no further vortexing or centrifugation.

Samples at various volume fractions are prepared as follows. A stock
suspension is centrifuged in a cylindrical cell for about 24 h at
1000~$g$, where $g$ is the acceleration of gravity.  to obtain a
random close packing (RCP) sediment, whose volume fraction, $\phi$,
is estimated to be $\phi = \varphi_\mathrm{rcp} \approx 0.67$
according to numerical results~\cite{refRCPpolydisperse}. It is
crucial to remark that this value of $\varphi_\mathrm{rcp}$ is
affected by a large uncertainty, since it depends on the details of
the particle size distribution, but also because the polymer layer
covering the particles may be slightly compressed during strong
centrifugation, implying that volume fraction is not accurately
known at this stage of the preparation.

Solvent is then added to, or removed from, the clear supernatant in
order to adjust the overall volume fraction to $\phi \approx 0.4$.
The suspension thus obtained is used as a mother batch from which
individual samples are prepared at the desired volume fraction by
adding or removing a known amount of solvent. All volume fractions
comparative to that of the initial batch are obtained with a
relative accuracy of $10^{-4}$, using an analytical balance and
literature values of the particle and solvent densities ($1190$
$\mathrm{kg\,m^{-3}}$ and $930$ $\mathrm{kg\,m^{-3}}$,
respectively)~\cite{Chaikin2}.

Although the {\it relative} volume fractions within our experimental
data are known with high accuracy, the {\it absolute} scale of
$\phi$ is far more difficult to estimate precisely, as there is no
direct way to measure $\phi$ which is not prone to uncertainty. We
discuss this issue in more detail in Sec.~\ref{tokuyama} below.

\subsection{Dynamic light scattering setup}

We use dynamic light scattering to probe the dynamics of our
concentrated colloidal hard spheres. In a DLS experiment one
measures the autocorrelation function of the temporal fluctuations
of the intensity scattered at an angle $\theta$~\cite{Berne1976}.
This allows the dynamics to be probed on a length scale $\sim
2\pi/q$, where $q = 2k_0\sin(\theta/2)$ is the magnitude of the
scattering vector, with $k_0$ the wave vector (in the solvent) of
the incident light, usually a laser beam.

Dynamic light scattering experiments are performed on samples
thermostated at $T = 27\pm0.1~ ^{\circ}\mathrm{C}$, using both a
commercial goniometer and hardware correlator (Brookhaven
BI9000-AT), to access the dynamics on time scales shorter than 10 s,
and a home-built, CCD-based apparatus to measure slower dynamics.
The CCD apparatus is described in Ref.~\cite{ElMasriJPCM2005}, and
so we simply recall here its main features. The source is a solid
state laser with \textit{in vacuo} wavelength $532.5$ nm and maximum
power 150 mW. The sample is held in a temperature-controlled copper
cylinder, with small apertures to let the incoming beam in and out
and to collect the scattered light. The beam is collimated, with a
diameter of 0.8 mm. All data reported here are obtained at a
scattering angle $\theta = 90$ degrees, corresponding to $q = 25.08$
$\mu \mathrm{m}^{-1}$ and $qR = 3.25$, slightly smaller than the
interparticle peak in the static structure factor. The collection
optics images a cylindrical portion of the scattering volume onto
the CCD detector, the diameter and the length of the cylinder being
approximately 0.8 and 2 mm, respectively.

We process the CCD data using the Time Resolved Correlation (TRC)
method~\cite{LucaJPCM2003}, which allows us to characterize
precisely both equilibrium and time-varying dynamics. The degree of
correlation, $c_I$, between pairs of images of the speckle pattern
scattered by the sample at time $\tw$ and $\tw+\tau$ is calculated
according to
\begin{equation}
c_I(\tw,\tau) = \frac{\left < I_p(\tw)I_p(\tw+\tau)\right >_p}{\left
< I_p(\tw)\right
>_p\left < I_p(\tw+\tau)\right >_p}-1 \, ,
\end{equation}
where $I_p$ is the intensity at pixel $p$ and the average is over
all CCD pixels. Data are corrected for the uneven distribution of
the scattered intensity on the detector as explained in
Ref.~\cite{DuriPRE2005}. For non-stationary dynamics (e.g. during
sample aging) the two-time intensity correlation function,
$g_2(\tw,\tw+\tau)-1$, is obtained by averaging $c_I(\tw,\tau)$ over
$\tw$, choosing a time window short enough for the dynamics not to
evolve significantly. For age-independent dynamics (e.g. when
equilibrium is reached), time invariance is fulfilled and
$g_2(\tw,\tw+\tau)-1$ reduces to the usual intensity correlation
function $g_2(\tau)-1$.

In order to obtain the full correlation function at all relevant
time delays, we first use the CCD apparatus to measure the dynamics
for time delays $\tau \geq 0.1 ~\mathrm{s}$. By applying the TRC
method, we monitor the evolution of the dynamics, until an
age-independent state is reached. The sample is then transferred to
the goniometer setup, where the fast dynamics ($\tau \leq 10$ s) are
measured using a point-like detector and averaging the correlation
function over time. The intensity correlation functions measured in
the two setups are then merged by multiplying the CCD data by a
constant, so that the two sets of data overlap in the range $0.1~
\mathrm{s} \leq \tau \leq 10 ~ \mathrm{s}$. Finally, the full
intensity correlation function is rescaled so that $g_2(\tau)-1
\rightarrow 1$ for $\tau \rightarrow 0$. Although in this paper we
focus on the slow dynamics and hence show only the CCD data, we
point out that the goniometer measurements are still necessary in
order to properly normalize the intensity correlation function. For
$\varphi \gtrsim 0.55$, we find that the dynamics are too slow to
obtain properly averaged data when only a time average is performed.
We thus adopt the so-called brute force method~\cite{XuePRA1992} for
the goniometer measurements: a large number of intensity correlation
functions are collected (typically 200 to 500), the sample being
turned in between two measurements in order to probe statistically
independent configurations. Additionally, the collection optics are
designed in such a way that about 10 coherence areas, or speckles,
are collected by the detector, further enhancing the statistics.

\subsection{Coherent vs. incoherent scattering}
\label{coherent-or-not}

The intensity correlation function is related to $f(\tw,\tw+\tau)$,
the (two-time) intermediate scattering function (ISF) or dynamic
structure factor, by the Siegert relation~\cite{Goodman1975}: $f =
\beta^{-1} \sqrt{g_2-1}$, where $\beta \leq 1$ is a constant that
depends on the collection optics. The ISF quantifies the evolution
of the particle position via
\begin{equation}
f(\tw,\tw+\tau) = \left\langle \frac{1}{N}
\sum_{j=1}^N  \sum_{k=1}^N
e^{i
{\bf q} \cdot \left [{\bf r}_j(\tw+\tau)-{\bf r}_k(\tw)\right ]}
\right\rangle. \label{fs}
\end{equation}
Here, the double sum is over $N$ particles, ${\bf r}_j(t)$ is the
position of particle $j$ at time $t$, $\mathbf{q}$ is the scattering
vector and brackets indicate the same average as for $g_2$. Note
that in general DLS measurements provide information on the
collective (or coherent) ISF, as indicated by the double sum over
$j$ and $k$ in Eq.~(\ref{fs}). For optically polydisperse samples,
however, the refractive index of the solvent may be tuned so that
the ISF reduces to its self (or incoherent) part, i.e. only terms
with $j=k$ contribute to the r.h.s. of Eq.~(\ref{fs}). Our sample is
optically polydisperse, because the PMMA core is polydisperse in
size and the grafted PHSA layer has a refractive index that is
different from that of the PMMA core, so that the average refractive
index of a particle varies with its size. Since the size
polydispersity is only moderate, one can safely assume that the
scattering power of a given particle is independent of the particle
position. Under this assumptions, the ISF can be written as the sum
of a ``full'' and a ``self'' (incoherent)
term~\cite{PuseyJPhysChem1982}. The contribution of the full term
vanishes if one matches the solvent refractive index to the mean
refractive index of the particles. In the experiments performed to
calibrate the absolute volume fraction by measuring the $\varphi$
dependence of the short-time dynamics for $\varphi \lesssim 0.2$
(see Sec.~\ref{tokuyama} and Fig.~\ref{toku} below), we carefully
adjust the refractive index of the solvent so as to fulfill this
condition. Additionally, these measurements are performed at $qR$ =
4.0, for which $S(q) \approx 1$, where the only contribution to the
ISF is the self part even if the average index is not
matched~\cite{PuseyJPhysChem1982}. Thus, in these runs we probe
uniquely the self part of the ISF, as required for applying the
calibration method described below.

In all the other measurements reported in this paper and in
Ref.~\cite{brambilla}, we deliberately increase by a small amount
the optical mismatch between the particles and the solvent, in order
to increase the intensity signal to be detected by the CCD camera.
By comparing the mean intensity for the slightly mismatched sample
to that measured for the best match, we find that the relative
weight of the ``full'' and ``self'' terms of the ISF are 35\% and
65\% respectively. The ``full'' term itself is a combination of the
self and distinct parts of the ISF, the relative weight of the self
part being of order $1/S(q)$~\cite{PuseyJPhysChem1982}. At the
scattering vector where the measurements are performed ($qR=3.25$),
$S(q)\lesssim 2$ for all studied volume fractions. Therefore, the
relative weight of the self term in the measured ISF is $\approx
80\%$ or more, and we conclude that
our DLS experiments probe essentially the self-part
of the ISF, namely:
\begin{equation}
f(\tw,\tw+\tau) \simeq  \left\langle \frac{1}{N}
\sum_{j=1}^N
e^{i
{\bf q} \cdot \left [{\bf r}_j(\tw+\tau)-{\bf r}_j(\tw)\right ]}
\right\rangle. \label{fself}
\end{equation}

\subsection{Determination of the volume fraction}
\label{tokuyama}

As mentioned above, the absolute value of the volume fraction after
preparation of a new sample is not known precisely, whereas the
uncertainty on the relative values of $\phi$ for a series of samples
obtained by dilution can be reduced to $10^{-4}$ or less. In most
works on nearly monodisperse colloidal hard spheres, the absolute
scale of the volume fraction is obtained by preparing samples in the
crystal-fluid coexistence region and by setting $\phi$ such that the
experimental melting and freezing fractions coincide with those
predicted theoretically~\cite{PuseyNature1986,vanmegen}. It is
important to remark that, unless the sample is perfectly
monodisperse, this method suffers from some uncertainty, since
numerical work shows that the volume fraction at freezing depends on
polydispersity. For example, in Ref.~\cite{vanmegen} the authors use
nominal values of $\phi$ that are obtained by comparison to the
freezing point of a monodisperse suspension, but they warn that the
true values of $\varphi$ could be up to a factor of 1.04 higher than
the nominal ones, due to a sample polydispersity $\sigma \approx
6~\%$. More generally, a relative error of about $5~\%$ is unavoidable
in experiments on hard sphere colloids.

\begin{figure}
\psfig{file=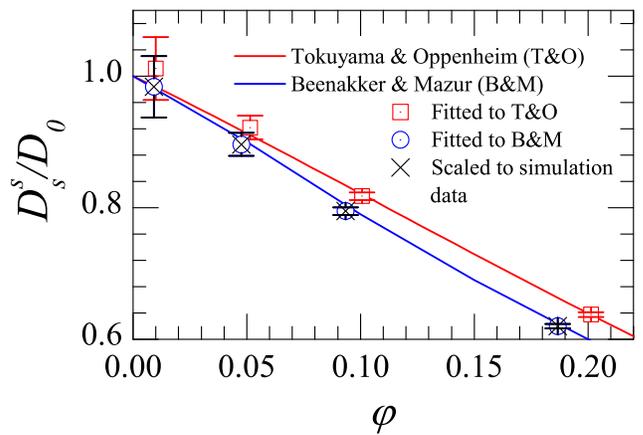,width=8.5cm} \caption{\label{toku}
(Color online) Volume fraction dependence of the dimensionless short
time self-diffusion coefficient. Open squares: experimental data
with the absolute volume fraction scaled to fit the
Tokuyama-Oppenheim prediction~\cite{Tokuyama}. Open circles: same
data scaled to the Beenakker-Mazur prediction~\cite{Beenakker}.
Crosses: same data scaled to maximize the overlap between the
experimental and numerical results for the structural relaxation
time, $\ta$, in the glassy regime $\phi \ge 0.55$. The scaling
factor for absolute $\phi$ between squares and circles is 1.07.}
\end{figure}

This calibration
procedure can not be applied to our more polydisperse suspensions,
since they do not crystallize. Instead, we calibrate the absolute
$\varphi$ by comparing the volume fraction dependence of the
short-time self-diffusion coefficient, $D_s$, to theoretical
predictions by Tokuyama and Oppenheim~\cite{Tokuyama} and Beenakker
and Mazur~\cite{Beenakker}. In practice, we start from an initial
guess of the absolute volume fractions, $\varphi_{\rm guess}$, and
measure $D_s$ \textit{vs.} $\varphi_{\rm guess}$ in the range
$\varphi_{\rm guess} \lesssim 0.2$ by fitting the initial decay of
the ISF to $f(q,\tau \to 0) \simeq \exp(-q^2D_s\tau)$. As explained in
Sec.~\ref{coherent-or-not}, we measure $f$ at $qR = 4$ and under the
best index matching conditions, to make sure that only the self part
of the ISF is probed. We then fit $D_s(\varphi_{\rm guess})$ to the
prediction for $D_s(\varphi)$ of either Ref.~\cite{Tokuyama} or
Ref.~\cite{Beenakker}, setting $\varphi = b \varphi_{\rm guess}$. There
are two
fitting parameters in this procedure: $b$ is the scale factor
for the absolute volume
fractions, and $D_0 \equiv \lim_{\phi \to 0}
D_s(\varphi)$ the self-diffusion coefficient in the limit of infinite
dilution.

The results of the fit are shown in Fig.~\ref{toku}
together with the theoretical curves. The data agree well with both
theoretical predictions, although the fit to the Beenakker-Mazur
expression is slightly better. The ratio of the scaling factors $b$
obtained by fitting the data to Ref.~\cite{Tokuyama} or
Ref.~\cite{Beenakker}, respectively, is 1.07. Thus, the spread in
the estimate of the absolute volume fractions is of order 7\%,
comparable to that in Ref.~\cite{vanmegen}. Interestingly, the
present  method was used in Ref.~\cite{SegrePRE1995} for a sample where
calibration using the experimental freezing point was simultaneously
possible, yielding fully compatible results.

As a final, alternative procedure to calibrate the absolute volume
fraction, we compare our data for the equilibrium relaxation times
described in more detail in Sec.~\ref{SEC:ANALYSIS} below, to the
results of Monte Carlo simulations for a three-dimensional binary
mixture of hard spheres~\cite{brambilla,longtom}. We find that it is
possible to make data in the glassy regime coincide over about 5
decades of relaxation times by adjusting the experimental scale for
$\phi$ by a factor very close to that required to fit $D_s$ to the
Beenakker-Mazur theory (see Fig.~\ref{toku}). In this paper, as
in Ref.~\cite{brambilla}, we adopt for convenience
the scaling factor required to
collapse the experimental and numerical relaxation times in the
glassy regime (crosses in Fig.~\ref{toku}).

This section shows that two distinct indicators obtained by
theoretical and numerical work can be used to adjust absolute volume
fractions in moderately polydisperse colloidal hard spheres,
yielding results that are consistent within a confidence interval of
about $\Delta \phi / \phi \approx 7~\%$, an uncertainty comparable
to that plaguing adjustments onto the static phase diagram due to
polydispersity effects. This uncertainty should be kept in mind when
absolute numbers for critical volume fractions are compared between
different works using different particles, different solvents,
different techniques, and different calibrations for
the volume fraction.

\subsection{Data analysis}
\label{treat}

\begin{figure}
\psfig{file=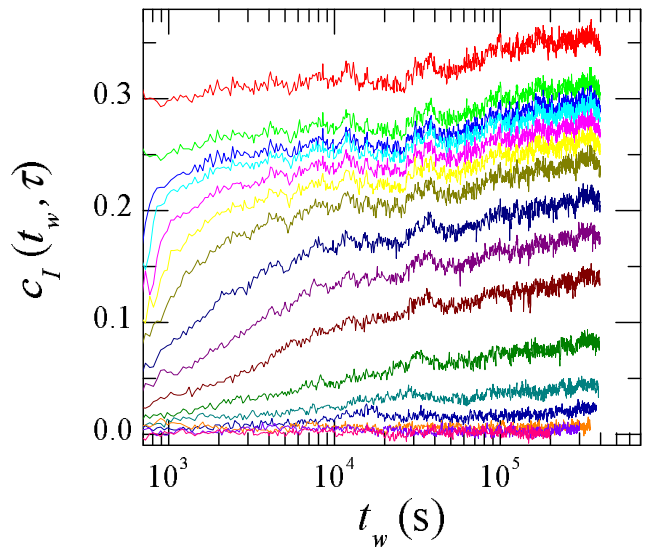,width=8.5cm}
 \caption{(Color online) Degree of correlation $c_I(\tw,\tau)$
measured for a sample with a volume fraction
 $\varphi = 0.5970$. From top to bottom, the time delay $\tau$
is 0, 1, 10, 20, 50, 100, 200, 500, 1k, 2k, 5k, 10k, 20k, 50k, 100k,
and 200k s.
 Note the logarithmic time axis, to better appreciate the initial
aging regime.}
 \label{Fig:cIdiluted}
\end{figure}

\begin{figure}
\psfig{file=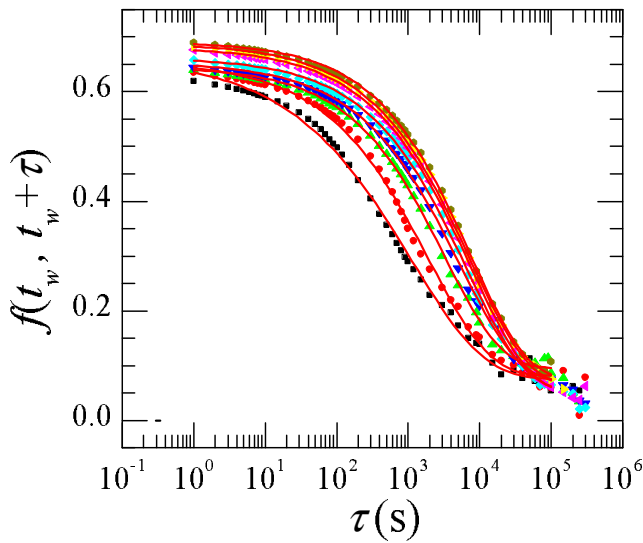,width=8.5cm}
 \caption{(Color online) Symbols: two-time intermediate scattering
functions (ISFs) $f(\tw,\tw+\tau)$ obtained from the degree of
correlation shown in Fig.~\ref{Fig:cIdiluted} ($\varphi=0.5970$).
From left to right, $\tw$ =1k, 2k, 5k, 10k, 50k, 100k, 200k, and
300k s. The solid lines are stretched exponential fits to the data
according to Eq.~(\ref{EqFitISF}).}
 \label{Fig:gIdiluted}
\end{figure}

As an example of the TRC data obtained in our experiments and their
processing, we show in Figs~\ref{Fig:cIdiluted} and
\ref{Fig:gIdiluted} the degree of correlation $c_I$ and the final
relaxation of the age-dependent ISF, respectively, for a sample at
$\varphi = 0.5970$. This is the densest system for which we were able to
collect data at thermal equilibrium.

The degree of correlation initially grows with
$\tw$, implying that the change in sample configuration over a fixed
time lag $\tau$ becomes progressively smaller, i.e. that the
dynamics slows down. After about $10^4-10^5$ s the evolution of the
dynamics essentially stops and a nearly stationary state is reached.
Figure~\ref{Fig:gIdiluted} shows the two-time ISF obtained by
averaging the $c_I$ data of Fig.~\ref{Fig:cIdiluted} over time
windows of duration 200 s (for the younger ages) to 1000~s (for the
oldest samples), for various ages $\tw$ (for clarity, the fast
dynamics measured with the goniometer apparatus is not shown). The
lines are stretched exponential fits to the final decay of the ISF:
\begin{equation}
f(\tw,\tw+\tau) = A(\tw)\exp \left \{ -\left [\tau/\taua(\tw)\right
]^{p\,(\tw)}\right \} + B\, , \label{EqFitISF}
\end{equation}
where $A$ is the height of the plateau preceding the final, or
$\alpha$, relaxation, $\taua$ the $\alpha$ relaxation time, $p$ the
stretching exponent and $B\geq 0$ a small, residual base line most
likely due to an imperfect correction of the effect of uneven
illumination, as discussed in Ref.~\cite{DuriPRE2005}. The aging
behavior observed for $c_I$ in Fig.~\ref{Fig:cIdiluted} is reflected
by the increase of $\taua$ with $\tw$. We find that for $\tw \geq 5
\times 10^4$ s the relaxation time does not evolve any more (see
also Fig.~\ref{Fig:TauTw} and the related discussion below), while
the height of the plateau still slightly increases, albeit extremely
slowly. We take $\tw = 5 \times 10^4$ as the end of the aging regime
and obtain the equilibrium value of $\ta$ as the average of the
fitted relaxation time in the stationary regime.

\section{Measuring long relaxation times}
\label{SEC:challenges}

We faced several challenges due to the extremely
long relaxation times of the system, the use of a CCD detector, and
the influence of gravity. In this section, we
describe some of the potential
artefacts and the difficulties associated with measurements on
nearly glassy samples, together with our solutions to overcome these
problems.

\subsection{Sample heating and convection.}

\begin{figure}
\psfig{file=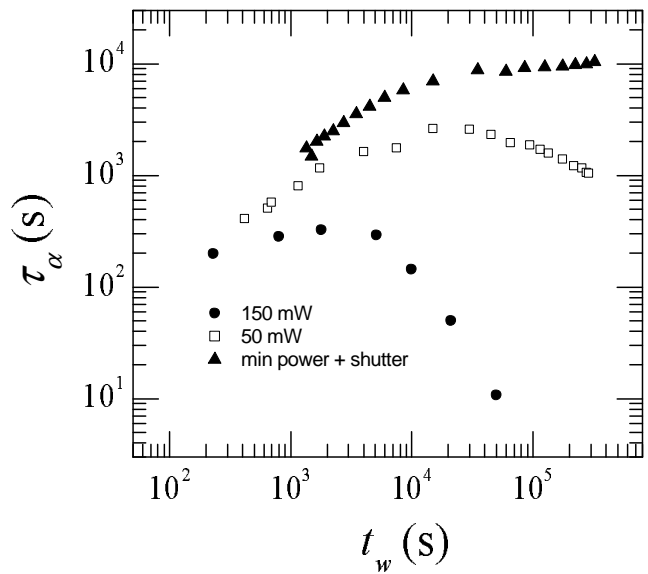,width=8.5cm}
 \caption{Age dependence of the $\alpha$ relaxation time for a sample at
 $\varphi=0.5970$. The curves are labeled by the laser power used
 for the measurement. The decrease of $\taua$ at large $\tw$ observed
at the higher laser powers is an artefact due to sample
 heating and convection, which is removed by reducing the power
and introducing a shutter.}
 \label{Fig:TauPhiLaser}
\end{figure}

The CCD detector is much
less sensitive than a traditional phototube or an avalanche
photodiode. Consequently, a larger laser power is required,
potentially leading to sample heating, if the particles or the
solvent absorb light at the wavelength of the source. Under normal
gravity conditions, local heating results in convective motion,
which can significantly alter the spontaneous dynamics of our
samples~\cite{bellon}.
Figure~\ref{Fig:TauPhiLaser} shows the age dependence of
the $\alpha$ relaxation time obtained by fitting the ISF to
Eq.~(\ref{EqFitISF}), for a sample at $\varphi=0.5970$. The
different curves are labeled by the power of the incident beam; when
the largest available power is used (150 mW, full circles), $\taua$
initially increases, as observed for many out-of-equilibrium
systems, but then drops sharply with increasing $\tw$. This
surprising ``rejuvenation'' effect is much less pronounced when a
reduced laser power is used (50 mW, open squares): the slowing down
of the dynamics persists over a longer time span and the decrease of
$\taua$ that eventually sets in is more modest. These observations
suggest that the acceleration of the dynamics, reported in one of
our previous papers~\cite{ElMasriJPCM2005}, is a non-linear effect
perturbing the equilibrium dynamics,
due to the onset of convective motion driven by sample heating.
Indeed, we
found that the solvent had became slightly yellowish since samples
were prepared, which was
responsible for increased absorption in the green region
of the visible spectrum, where our laser operates.

To avoid any
artefact due to convection, we have added a shutter to the setup, so
that the sample is illuminated only for 100 ms at each CCD
acquisition, rather than continuously. Since the typical acquisition
rate is 1 Hz or lower, this reduces the average power injected in
the sample by a factor of ten or more. Additionally, we perform our
experiments at the minimum laser power compatible with a good beam
stability (about 10 mW). Under these conditions, $\taua$ is found to
increase monotonically until a plateau is reached, with no
rejuvenation effect even after 2 days, as shown by the triangles in
Fig.~\ref{Fig:TauPhiLaser}. By changing the opening time of the
shutter, we have checked that the dynamics shows no dependency on
the average illuminating power, as long as the laser power is $\leq
10$ mW and the shutter time is $\leq 1$ s. As a final remark on this
spurious rejuvenation, we note that this phenomenon sets in very
slowly, especially when the laser power is not too high. This hints
at a subtle interplay between aging and an external drive, similarly
to the behavior of other soft glassy materials to which a (modest)
mechanical perturbation is applied during their
aging~\cite{CoussotPRL2002}. More experiments will be needed to
explore these analogies.

\subsection{Sedimentation}

Sedimentation is a potential additional source of
a spurious acceleration of
the dynamics, since it can act as a non-linear driving force.
This is  especially true for systems that relax very slowly and
are very sensitive to external forcing~\cite{sgr,2time}.
In the
imaging geometry of our CCD setup, sedimentation results in an
overall drift of the speckle pattern on the detector, which
contributes to the change of $I_p$ at any given pixel and thus to
the decay of $g_2$. We have implemented a correction scheme, which
will be described in detail in a forthcoming publication; here we
explain only the principles of the method. We use Image Correlation
Velocimetry (ICV)~\cite{TokumaruExpInFluids1995}, a technique
similar to Particle Imaging Velocimetry, to measure the drift of the
speckle pattern due to sedimentation for each pair of images taken
at time $\tw$ and $\tw + \tau$. The drift is obtained by calculating
the spatial crosscorrelation between the two speckle images. If the
relative displacement of the particles over the lag $\tau$ is
smaller than $1/q$, the lengthscale probed in a DLS experiment, the
speckle pattern is essentially frozen, except for its overall drift.
The second speckle image is then a shifted version of the first one;
consequently, the crosscorrelation function exhibits a marked peak
whose position corresponds to the average particle drift. Once the
drift is obtained, the second image is back-shifted
numerically~\cite{NobachExpFluids2005} by the same amount but in the
opposite direction, so as to compensate the physical drift. The
corrected $c_I$ is finally calculated between the first image and
the back-shifted version of the second one. The minimum shift that
can be reliably measured is a few hundredths of a pixel,
corresponding to about 50-100 nm in the sample. This is slightly
less than $2\pi/q = 250$ nm, the length scale over which the
relative motion is probed by DLS: thus, we are able to detect and
correct for an overall drift comparable to the relative displacement
of the particles. Note that our correction method fails when the
speckle pattern ``boils'' and changes faster than it drifts, since
no peak in the crosscorrelation function can be detected reliably.
However, this situations corresponds precisely to the case where the
internal, spontaneous dynamics dominates over sedimentation, making
the correction for the drift irrelevant.

\begin{figure}
\psfig{file=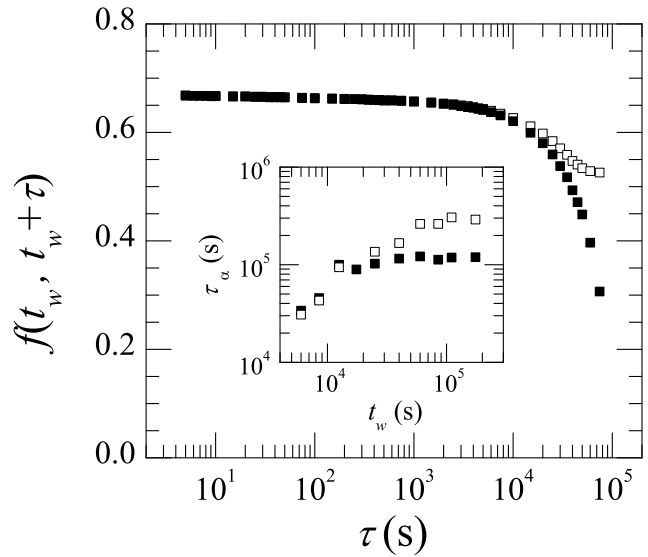,width=8.5cm}
 \caption{Main panel, filled squares: final relaxation of the ISF for
a concentrated sample at
 $\varphi=0.5990$ and $\tw = 10^5$ s. The open symbols are the same data
 after correcting for the contribution of the average sedimentation
 velocity to the decay of $f$. Inset: for the same $\varphi$, $\tw$
dependence of the
 $\alpha$ relaxation time for the raw ISFs and the corrected data.}
 \label{Fig:Sedimentation}
\end{figure}

Figure~\ref{Fig:Sedimentation} shows both the raw (solid squares)
and the corrected (open squares) $g_2-1$ for the most concentrated
sample that we have studied ($\phi = 0.5990$) at $\tw = 10^5$ s.
Sedimentation clearly affects the decay of $g_2$, which is about
three times faster for the raw data than for the correct ones. The
inset shows the large-age behavior of the relaxation time extracted
from raw (open squares) and corrected (solid squares) correlation
functions. Sedimentation leads to an apparent arrest of aging for
$\tw \geq 10^4$ s, while the corrected data indicate that the sample
barely equilibrates for $\tw \geq 10^5$, an age ten times larger.
Note that, although we do correct for the average sedimentation
velocity, we can not correct for velocity fluctuations stemming
from hydrodynamics interactions and possibly
changing the relative position of
the particles, thereby contributing to the loss of correlation of
the scattered light. Since velocity fluctuations are known to be
relevant in concentrated suspensions~\cite{NicolaiPhysFluids1995},
it is likely that sedimentation still accelerates to some extent the
decay of $g_2$. These data are therefore not included in the
analysis of equilibrium dynamics below.

It is interesting to compare the potential impact of sedimentation
on our results to that in other works. A recent set of
experiments~\cite{SimeonovaPRL2004} on concentrated PMMA hard
spheres has shown that aging is faster for off-buoyancy-matching
particles than for nearly-buoyancy-matched colloids. Although our
particles are not buoyancy-matched ($\Delta \rho =
259~\mathrm{kg}\,\mathrm{m}^{-3}$), they are smaller than those used
in the optical microscopy investigation of
Ref.~\cite{SimeonovaPRL2004}, enhancing the role of Brownian
diffusion with respect to the gravity-driven drift. The relevant
parameter is the inverse Peclet number \be \Lambda =
\frac{3k_\mathrm{B}T}{4 \pi g \Delta \rho R^4} \, , \ee defined as
the ratio of the gravitational length to the particle radius ($T$ is
the absolute temperature and $k_\mathrm{B}$ the Boltzmann's
constant). In this work, $\Lambda \approx 870$, much larger than
$\Lambda = 44.1$ for the ``normal gravity'' experiments of
Ref.~\cite{SimeonovaPRL2004} (for which $\Delta \rho =
300~\mathrm{kg}\,\mathrm{m}^{-3}$) and still greater than $\Lambda =
678$ for their ``reduced gravity'' measurements ($\Delta \rho =
20~\mathrm{kg}\,\mathrm{m}^{-3}$). Thus, our experiments are
comparable to typical ``reduced gravity'' microscopy investigations.
This illustrates the great advantage of using smaller particles to
mitigate gravity effects, due to the $R^{-4}$ dependence of
$\Lambda$. In order to further investigate the role of gravity in
the slow dynamics of colloidal hard spheres in the large $\Lambda$
regime, light scattering measurements on small particles that are
partially buoyancy-matched or microgravity experiments will be
needed.

As a final remark on potential artefacts, one may wonder whether
mechanical instabilities could be partially responsible for the
ultra-slow decay observed in Fig.~\ref{Fig:Sedimentation}, where the
relaxation time can be as large as 1.5 days. To test this
possibility, we have measured $g_2$ for a sample compressed at random close
packing and where very little dynamics is expected to occur.
Although $g_2$ does exhibit some decay, whose origin is
still unclear (either mechanical instability or some extremely slow
rearrangement of the packed particles, or the contributions of a
small number of ``rattlers''), we stress that this relaxation is
much slower than that observed for the most concentrated sample
studied in this work. Furthermore, the relaxation of the most
concentrated sample (Fig.~\ref{Fig:Sedimentation}) is a factor of 10
slower than the slowest equilibrium relaxation reported here and in
Ref.~\cite{brambilla}: we can therefore rule out any significant
artefact due to mechanical instability.

\subsection{Non-equilibrium aging effects}

\begin{figure}
\psfig{file=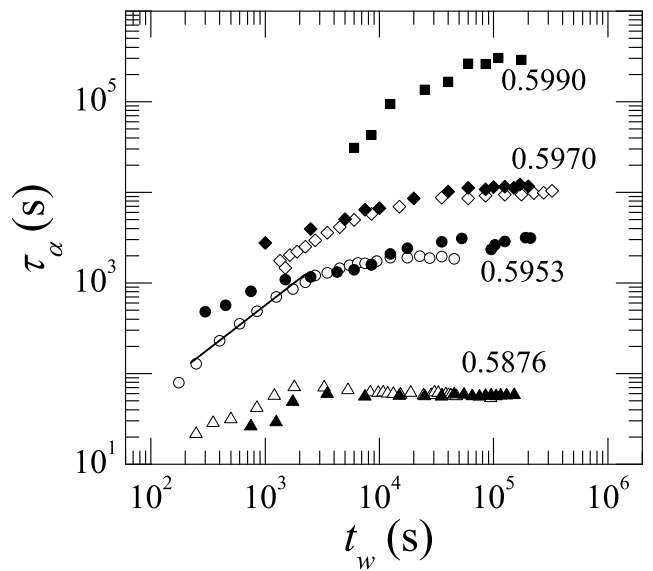,width=8.5cm}
 \caption{Age dependence of the $\alpha$ relaxation
time for samples at volume fractions $0.5876 \le \varphi \le
0.5990$, as indicated by the labels. Different symbol fillings
correspond to independent experiments  on the same sample. Large-$\tw$ data
for the two largest volume fractions are corrected for sedimentation
effects. The line has a slope of one, showing that the initial
growth of $\taua$ is in general close to linear.} \label{Fig:TauTw}
\end{figure}

Besides artefacts due to sample heating and sedimentation, the
dynamics of very concentrated samples become intrinsically difficult
to measure, because of aging and dynamical heterogeneity. Aging is a
quite general feature of glassy systems (see
e.g.~\cite{CipellettiJPCM2005} and references therein), where the
dynamics slows down with time, or sample ``age'', $\tw$, as the
system evolves towards equilibrium. In traditional DLS, the
intensity correlation function has to be extensively averaged over
time, making it impossible to capture the evolution of the dynamics
for non-stationary systems. By contrast, the TRC approach and other
multispeckle
methods~\cite{KirschJChemPhys1996,LucaRSI1999,ViasnoffRSI2002} allow
one to fully characterize time-evolving dynamics, because the time
average is replaced in part, or completely, by an average over the
slightly different $q$ vectors associated to distinct pixels of the
CCD.

Figure~\ref{Fig:TauTw} shows the age dependence of $\taua$ for
samples prepared at different volume fractions $\varphi \ge 0.5876$.
Symbols with the same shape but different filling correspond to
independent experiments at the same volume fraction. Data for the two
largest $\phi$ are corrected for sedimentation effects at the
largest $\tw$; in all other cases, corrected and raw data yield the
same results. Initially, $\taua$ increases with $\tw$; in this
regime, the growth of $\taua$ $vs.$ $\tw$ is generally close to
linear, as shown by the line $\taua \propto \tw$. For the most
diluted sample in Fig.~\ref{Fig:TauTw} ($\varphi = 0.5876$), the
initial aging regime is very short and is comparable to the time
needed by the sample to equilibrate at the temperature imposed by
the copper holder; consequently, no precise aging law can be
determined. After the initial regime, $\taua$ saturates and becomes
almost independent of time, suggesting that all samples reach
equilibrium, with the possible exception of the most concentrated
suspension ($\varphi=0.5990$, squares in Fig.~\ref{Fig:TauTw}), for
which the dynamics seem to keep slowing down over the full duration
of the experiment (more than 3 days), although at an increasingly
smaller rate. Note that the experiments are quite generally well
reproducible, with the exception of the earliest stages, where
presumably the dynamics is more influenced by the exact initial
configuration and thermalization.

\begin{figure}
\psfig{file=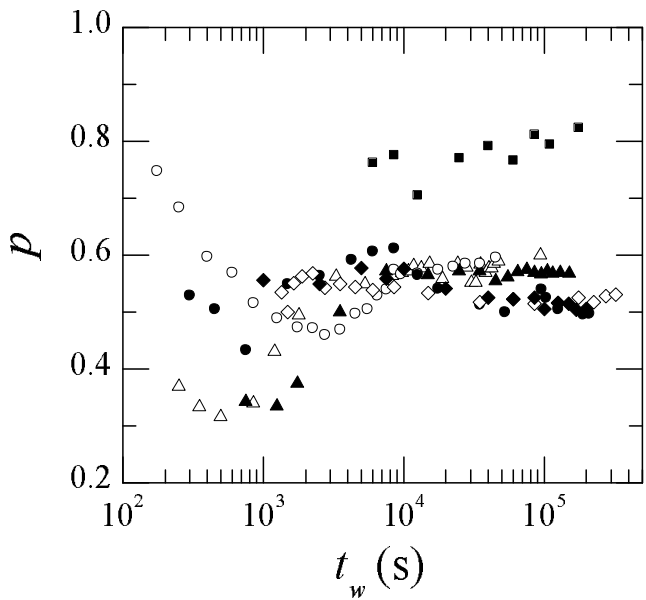,width=8.5cm}
 \caption{Age dependence of the stretching exponent $p$
obtained by fitting the final decay of the ISF to
 Eq.~(\ref{EqFitISF}).
 Same symbols as in Fig.~\ref{Fig:TauTw}.}
 \label{Fig:pTw}
\end{figure}

The time needed for reaching equilibrium, $t_{\mathrm{eq}}$, is
about 30 times greater than the (equilibrium) relaxation time for
the samples at the lowest volume fraction, $\varphi=0.5876$.
However, this values is certainly overestimated due to the
contribution of the thermalization time, of the order of $10^3$ s.
For the samples at $\varphi = 0.5953$ and $0.5970$, for which full
equilibration is reached and where the initial thermalization time
is negligible compared to all relevant time scales,
$t_{\mathrm{eq}}/\taua \approx 3-4$. This confirms the intuitive
notion that equilibrium is reached on the time scale of a few
structural relaxation times, i.e. after a few cage escape processes.
Note that we are able to show that there is indeed an equilibrium
regime even for samples whose $\alpha$ relaxation time is more than
7 orders of magnitude larger than that of very dilute suspensions.
This significantly extends the range of samples that do reach
equilibrium, compared to previous work~\cite{vanmegen}, where
samples with $\taua$ larger than $10^5$ times the Brownian decay
time were classified as aging samples, most likely due to technical
difficulties to access much larger relaxation timescales.

The time evolution of the stretching exponent $p(t_w)$ is shown in
fig.~\ref{Fig:pTw}. For sufficiently large $\tw$, $p$ saturates at
0.5-0.6, regardless of the volume fraction of the particles and the
history of the samples. This stretched exponential shape might be
related to sample polydispersity, since in previous experimental
work on similar but less polydisperse colloidal hard spheres
($\sigma \approx 6\%$) the ISF was shown to relax nearly
exponentially~\cite{vanmegen}. However, we point out that we find a
similar value in the simulations described below in
Sec.~\ref{simu}, for both $\sigma = 5.77\%$ ($p\approx 0.55$)
and $\sigma = 11.5\%$ ($p\approx 0.65$).
Interestingly, for the most concentrated sample $p$ remains well
above the asymptotic value measured for all other samples, further
suggesting that this sample has not fully relaxed during the
experimental time window. Note that here we never observe
``compressed'' exponential relaxations ($p > 1$), in contrast to the
data of Ref.~\cite{ElMasriJPCM2005}. We recall that for the latter
laser heating yielded convection, which was most likely responsible
for faster-than-exponential relaxations.

\subsection{Dynamic heterogeneity}

\begin{figure}
\psfig{file=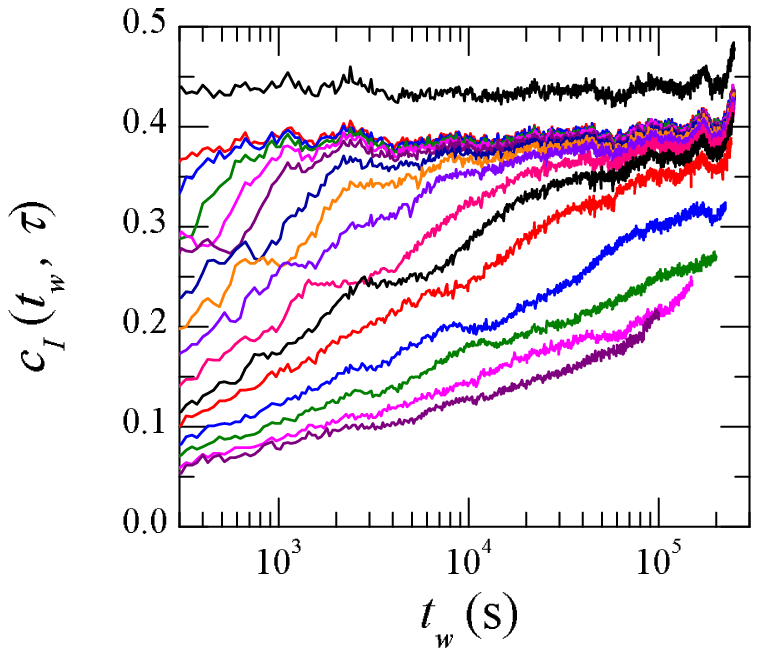,width=8.5cm}
 \caption{(Color online) Degree of correlation for the most
concentrated sample ($\varphi=0.5990$). Note the large
 fluctuations of $c_I$, especially for $t_{\mathrm{w}} \leq 5
\times 10^3$ sec, indicative of a temporally heterogeneous dynamics.
From top to bottom,
 $\tau$ is 0, 5, 10, 25, 50, 100, 250, 1k, 2.5k, 5k, 10k, 25k, 50k,
100k, and 150k sec.}
 \label{Fig:cI50000}
\end{figure}

\begin{figure}
\psfig{file=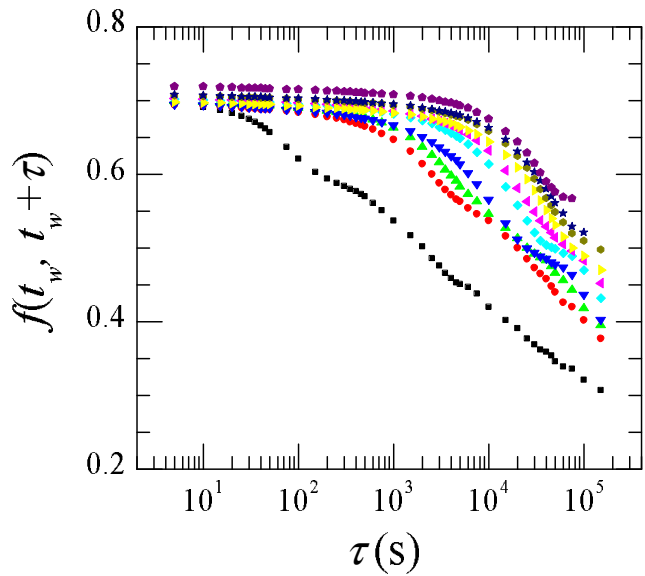,width=8.5cm}
 \caption{(Color online) Two-time ISFs obtained from the $c_I$
data shown in Fig.~\ref{Fig:cI50000} ($\varphi=0.5990$).
 From left to right, $\tw=500$, 5k, 7k,
 10k, 20k, 30k, 50k, 70k, 100k, and 170k s.
 Dynamical heterogeneity at early times results in noisy ISFs, whose
 shape significantly departs from stretched exponential.}
 \label{Fig:gI50000}
\end{figure}

A detailed discussion of dynamical heterogeneity in samples at
intermediate volume fractions ($0.20 < \varphi < 0.576$) has been
presented in Refs.~\cite{BerthierScience2005,brambilla,dalleferrier}.
Here, we
simply report that both the degree of correlation $c_I$ and the
age-dependent ISF become more erratic and exhibit significant
temporal fluctuations as $\varphi$ grows. An example is shown in
Fig.~\ref{Fig:cI50000} for $c_I$, and in Fig.~\ref{Fig:gI50000} for the
correspondent time-resolved ISFs. The data presented here are for
the most concentrated sample that we have studied, $\varphi =
0.5990$. Note that the fluctuations of $c_I$ are particularly
relevant at early ages, when the sample is presumably farther from
its equilibrium configuration. Because of these fluctuations, the
shape of $f(\tw,\tw+\tau)$ is less well defined, and the fits are
generally poorer. However, we stress that $\taua$ remains reasonably
well defined, since the decay of the ISF, although somehow
``bumpy'', is generally not too stretched, with the exception of the
earliest ages.

The observation that $c_I$ and $f$ are increasingly noisy as
$\varphi$ increases is consistent with the recently reported growth
of cooperatively rearranging regions in supercooled colloidal
suspensions approaching the glass
transition~\cite{WeeksScience2000,BerthierScience2005,brambilla,WeeksJPCM2007},
in analogy with the behavior of molecular glass
formers~\cite{EdigerReview,GlotzerReview}. Indeed, as the size of
the regions that rearrange cooperatively increases, the number of
such regions in the scattering volume decreases: the dynamics are
averaged over a smaller number of statistically independent objects,
leading to enhanced fluctuations~\cite{foampeter}. The fact that dynamical
fluctuations at high $\varphi$ are significant suggests that
cooperatively rearranging regions may extend over a sizeable
fraction of the scattering volume. This would imply a correlation
length of the dynamics of the order of several hundreds of particle
diameters, much larger than the $\sim 10$ particles reported in
previous
works~\cite{WeeksScience2000,BerthierScience2005,brambilla,WeeksJPCM2007}
at lower $\varphi$, but comparable to recent findings in
polydisperse colloids near random close
packing~\cite{BallestaNatPhys2008}. More experiments will be
required to confirm these intriguing preliminary results.

\section{Analysis of equilibrium dynamics}
\label{SEC:ANALYSIS}

\subsection{Equilibrium relaxation times}

The first two sections were meant to convince the reader that
it is possible to access the equilibrium
dynamics of our colloidal hard sphere
sample over a broad range of timescales for several
well-controlled volume fractions. We now turn to the analysis
of these results, which were briefly presented
in a shorter paper~\cite{brambilla}.

\begin{figure}
\psfig{file=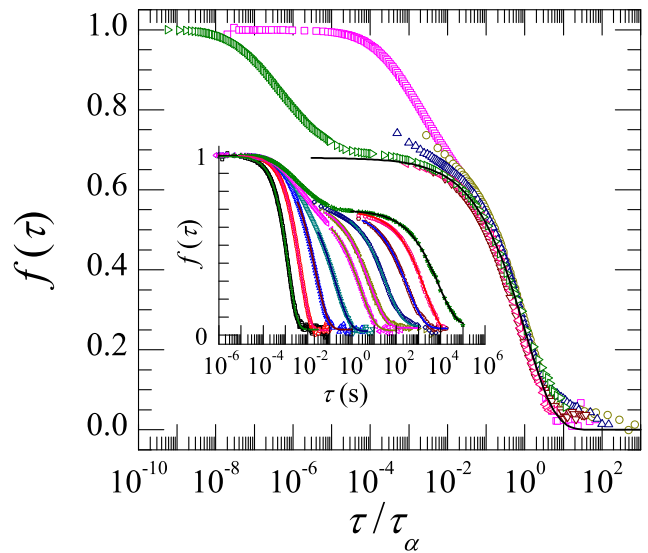,width=8.5cm}
\caption{\label{figfscaling} (Color online) Inset: representative
ISFs for equilibrated samples at volume fractions 0.0480, 0.3096,
0.4967, 0.5555, 0.5772, 0.5818, 0.5852, 0.5916, 0.5957, 0.5970, from
left to right. The lines are stretched exponential fits to the final
decay of $f$. Main figure: scaling of the $\alpha$ relaxation of the
ISFs for $\phi \geq 0.4967$ when using the reduced variable
$\tau/\tau_\alpha$. The line is the function
$0.68\exp[-(\tau/\tau_\alpha)^{0.558}]$.}
\end{figure}

In the inset of Fig.~\ref{figfscaling} we show the time decay of the
ISF at selected volume fractions from a dilute system at $\phi
\approx 0.05$, up to very large volume fractions, $\phi=0.597$ (same
data as in Ref~\cite{brambilla}). Our data cover a broad range of 11
decades in time, and we follow the slowing down of the equilibrium
dynamics over about 7 decades in relaxation times.

In agreement with previous work~\cite{pusey2}, we find that time
correlation functions decay exponentially when volume fraction is
moderate, with a time constant that increases weakly with $\phi$.
When $\phi$ is increased above some `onset' volume fraction, $\phi
\approx 0.5$, the relaxation becomes strongly non-exponential, with
a two-step decay that become more pronounced as $\phi$ continues to
grow, signalling the arrest of particle motion for intermediate
timescales. While the short-time decay is relatively unaffected by
the increase of $\phi$, the characteristic time of the second decay
corresponding to the structural relaxation of the fluid increases
rapidly over a small range of volume fraction. For $\phi > 0.597$,
we were not able to collect data for which stationary behavior and
thermal equilibrium could be unambiguously established, and this
volume fraction marks thus the location of the `experimental
colloidal glass transition', $\phi_g \approx 0.6$, by analogy with
the temperature scale $T_g$ in thermal glasses. We do not attach any
particular significance to $\phi_g$, because it obviously depends on
the details of the experimental set-up and of the sample.

As described in Sec.~\ref{treat}, we fit the long-time decay of the
time correlation functions to a stretched exponential form,
Eq.~(\ref{EqFitISF}). The resulting fits are shown in the inset of
Fig.~\ref{figfscaling} as continuous lines. They describe the data
very well and yield quantitative confirmation of the above
qualitative remarks. We find that the amplitude $A$ of the decay is
$A \approx 1$ at low $\phi$ when relaxation is mono-exponential, so
that $p \approx 1$ in that regime. When $\phi$ increases, we find
that $A$ drops to $A \approx 0.7$, signalling a two-step process. At
the same time, the stretching exponent $p$ decreases and stabilizes
around $p \approx 0.56$, indicating that structural relaxation
occurs through a broad distribution of timescales.

The fact that $p$ is only weakly dependent on the volume fraction
suggests that the structural decay of the ISFs should collapse when
plotted against rescaled time, in the spirit of the `time
temperature superposition principle' frequently observed in
molecular glass-formers. We show such a scaling in the main plot of
Fig.~\ref{figfscaling}, where the ISFs for $\phi \geq 0.4967$ are
plotted as a function of $\tau/\ta$, with $\ta$ the structural
relaxation time issued from the stretched exponential fit to the
data. The data collapse reasonably well onto a single master curve,
confirming that the time volume fraction superposition principle
holds for colloidal hard spheres, as observed in~\cite{vanmegen}.

\begin{figure}
\psfig{file=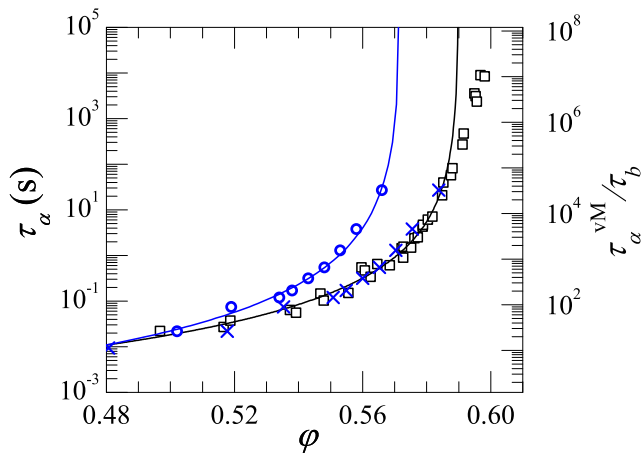,width=8.5cm}
\caption{\label{figtauexperiments} (Color online) Open squares, left
axis: volume fraction dependence of the $\alpha$ relaxation
timescale for equilibrated samples. The black line is an MCT fit to
the data, Eq.~(\ref{mct}). Open circles, right axis: $\alpha$
relaxation timescale measured by van Megen et al. in
\cite{vanmegen}, in units of the Brownian time, $\tau_b =
R^^2/(6D_0)$. The blue line is an MCT fit to the data of
\cite{vanmegen}. Crosses: same data as the circles, plotted as a
function of volume fraction scaled by a factor of 1.031, a value
smaller than the uncertainty on $\phi$ stated in \cite{vanmegen}.
The scaled data satisfactorily overlap with the measurement of this work.}
\end{figure}

We present in Fig.~\ref{figtauexperiments} the evolution of the
equilibrium relaxation time $\ta(\phi)$ extracted from the fits of
the ISFs. For the sake of clarity, only data for $\phi \ge 0.48$ are
shown: data at all volume fractions were reported in
Ref.~\cite{brambilla}. While $\ta$ increases by about 1 decade from
$\phi=0.05$ to $\phi=0.5$ (data not shown), the structural
relaxation slows down dramatically by about 6 decades when $\phi$
approaches the largest volume fraction considered in this work.
Following previous analysis~\cite{vanmegen}, we fit $\ta$ to an
algebraic divergence, as in Eq.~(\ref{mct}), using $\phi_c$ and
$\gamma$ as adjustable parameters. Such a power law, predicted by
MCT, cannot account for our data over the whole range of volume
fraction, and the range of volume fractions fitted by
Eq.~(\ref{mct}) must be chosen with great care. As is also found in
molecular glass-formers, we find that the power law Eq.~(\ref{mct})
describes a window of relaxation times of about three decades,
immediately following the onset of glassy dynamics, as shown in
Fig.~\ref{figtauexperiments}. The values of the exponent, $\gamma =
2.5 \pm 0.1$, and of the critical volume fraction, $\phi_c = 0.590
\pm 0.005$, depend on the precise range of volume fraction fitted,
but they agree well with previous work~\cite{vanmegen}, and
theoretical and numerical analysis~\cite{brambilla,longtom}, as we will
further discuss in Sec.~\ref{SUBSEC:comparison}. If we attempt to
include the largest volume fractions in our fit, we find that
$\gamma$ takes unphysically large values, up to $\gamma \approx 6$.
We interpret this finding as the sign that the growth of $\ta$ is
not appropriately described by Eq.~(\ref{mct}) at large volume
fraction.

A more direct indication can be observed in Figs.~\ref{figfscaling}
and \ref{figtauexperiments}, since we were able to collect
{\it equilibrium} data at volume fractions $\phi$ {\it above} the fitted value
of the critical volume fraction $\phi_c$. Therefore, the dynamic
singularity implied by Eq.~(\ref{mct}) is not observed in our
sample, and ergodic behavior can be detected for $\phi>\phi_c$. Just
as is generically found in thermal glasses, we conclude therefore
that the singularity predicted by mode-coupling theory is avoided in
our experimental colloidal system. Although suggested by theory and
computer simulations, such an observation was not reported in
experimental work before.

As discussed in detail in Ref.~\cite{brambilla}, we find that an
exponential divergence of $\ta$ accounts for our data very well, as
in Eq.~(\ref{vft}). We argued in Ref.~\cite{brambilla} that the best
fit of our data was obtained for $C = 9.8 \times 10^{-3}$,
$\tau_{\infty} = 6.5\times 10^{-3}$ sec, $\delta = 2.0 \pm 0.2 $ and
$\phi_0 = 0.637 \pm 0.002$, although imposing $\delta = 1$, as in
free volume predictions, yields an acceptable fit and a smaller
critical volume fraction, $\phi_0(\delta=1) = 0.614 \pm 0.002$.

Interestingly, the non-trivial exponent $\delta \approx 2$ is also
supported by the numerical data for the polydisperse system of
quasi-hard spheres studied in Sec.~\ref{simu} (see
Fig.~\ref{tausimu} below), and by Monte Carlo results for a binary
mixture of hard spheres~\cite{brambilla,longtom} (also shown in
Fig.~\ref{tausimu}). Additionally, a recent scaling analysis of the
glass transition occurring in systems of soft repulsive particles
yields $\delta = 2.2 \pm 0.2$ for the hard sphere limit~\cite{longtom,tom},
a value compatible with that obtained here from the fit of
Eq.~(\ref{vft}). Thus, there is mounting evidence from both
simulations and experiments that a simple VFT law, Eq.~(\ref{vft})
with $\delta=1$, is not the most accurate description of the
dynamics of hard spheres.

Overall, we find that while the onset of dynamical slowing can be
described by an MCT divergence at a critical volume fraction
$\phi_c$, upon further compression a crossover from an algebraic to
an exponential divergence at a much larger volume fraction $\phi_0$
is observed, showing that the apparent singularity at $\phi_c$ does
not correspond to a genuine `colloidal glass transition'. Instead
the system enters a regime where dynamics is `activated', in the
sense that $\ta$ increases exponentially fast with the distance
to $\phi_0$. These observations
therefore suggest that the glass transition scenario in colloidal
hard spheres strongly resembles the one observed in molecular
glass-forming materials.

\subsection{Polydispersity effects}
\label{simu}

A central finding from our experiments is the absence of a genuine
dynamic singularity occuring at the fitted mode-coupling singularity
$\phi_c$. Since polydispersity influences the dynamics of colloidal
hard spheres, we must ask how our findings may change for samples
with a different polydispersity.

In Ref.~\cite{vanmegen} a sample less polydisperse than ours
($\sigma \approx 6\%$ as opposed to $\sigma \approx 10\%$) was
analyzed using similar light scattering techniques, but no deviation
from an algebraic singularity was reported. Additionally,
experiments performed on samples at different polydispersities
showed that increasing the polydispersity at constant volume
fraction produces an acceleration of the dynamics~\cite{poly}. These
observations suggest a possible explanation for our findings. One
could argue that the observed ergodic behavior in our sample at large
$\phi$ stems from the fact that the glass state above the critical
volume fraction $\phi_c$ determined in Ref.~\cite{vanmegen} is
`melted' by polydispersity effects. In other words, polydispersity
could push $\phi_c$ to a much larger value so that our fluid states
could all be observed below the ``true'' $\phi_c$ characterizing our
sample. This view is apparently supported by numerical
analysis~\cite{refRCPpolydisperse}, which showed that volume
fractions for the location of random close packing could be shifted
from $\phi_{\rm rcp}=0.64$ to $\phi_{\rm rcp}=0.67$,
from monodisperse to 10~\%
polydisperse samples, respectively, thus suggesting a similarly
large shift in the MCT critical volume fraction.

This string of arguments is immediately contradicted by the fact
that we have self-consistently determined $\phi_c$ for our own
sample, and that we do observe ergodic behavior for $\phi$ which are
larger than this critical volume fraction. It remains thus to
establish that polydispersity does not qualitatively affect the
dynamics in such a drastic manner that the results reported in this
article do not shed light on the behavior of less polydisperse
samples.

To definitely settle this issue, we performed computer simulations
of a quasi-hard sphere system at different
polydispersities~\cite{DjamelPhD}. Following previous
work~\cite{puertas}, we study an assembly of point particles
interacting via a purely repulsive potential: \be V(r_{ij}) =
\epsilon \left( \frac{r_{ij}}{2 R_{ij}} \right)^{36}, \label{h36}
\ee where $r_{ij}$ is the distance between particles $i$ and $j$,
and $R_{ij} = (R_i + R_j)/2$ with $R_i$ the radius of particle $i$.
The prefactor $\epsilon$ is an energy scale. The interaction
potential is very steep, and should therefore constitute a good
approximation of the hard sphere potential. We set the
polydispersity by drawing the particle radii from a flat
distribution: $R_i \in [ R-\delta R/2 , R+\delta R/2]$, so that the
average diameter is $\overline{R_i} = R$ and the polydispersity is
$\sigma = \sqrt{\overline{R_i^2} - R^2}/R = \delta R /(\sqrt{12}R)$.
We have studied the system for two values of polydispersity, $\delta
R/R = 0.2$ and 0.4, leading to polydispersities $\sigma \simeq
5.77~\%$ and $\sigma \simeq 11.5~\%$, which compare well with the
colloidal samples studied in Ref.~\cite{vanmegen} and in the present
study, respectively.

We follow previous work dedicated to the Hamiltonian (\ref{h36}) in
the context of hard spheres. We solve Newton's equations for $N$
particles of mass $m$ inside a three-dimensional simulation box of
size $L$ with periodic boundary conditions~\cite{allen}. For the inverse power
law potential in Eq.~(\ref{h36}), density and temperature can be
combined in a unique control parameter, $\Gamma = \phi T^{-1/12}$,
so we fix the temperature and energy scales, $k_B T = \epsilon =
1/3$, and vary the system size to change the volume fraction, $\phi
= 4\pi N R^3 [1+(\delta R/4R)^2 ]/(3L^3)$. Since we deal with
soft spheres, it is of course
not possible to compare absolute values for critical
volume fractions with results obtained in the true
hard sphere limit. Timescales are expressed in units of $2R
\sqrt{m/\epsilon}$. We choose the self-intermediate scattering
function $F_s(q,t)$ as a dynamic observable, and work at a single
wavevector, $q R = 3.9$, close to the first diffraction peak.

\begin{figure}
\psfig{file=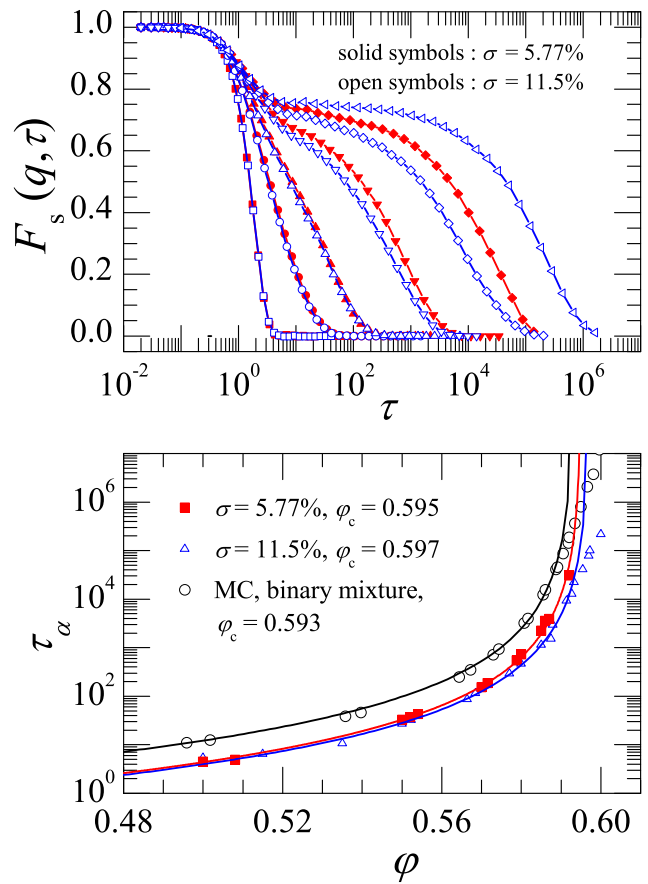,width=8.5cm}
 \caption{\label{tausimu} (Color online) Top panel:
self-intermediate scattering function from numerical simulations of
a quasi-hard sphere system at two different values of the
polydispersity, $\sigma$, and various volume fractions. For the less
polydisperse system, $\phi = 0.053$, 0.50. 0.55, 0.58 and 0.592,
from left to right. For the more polydisperse system, $\phi =
0.074$, 0.50, 0.55, 0.58, 0.5914 and 0.60, from left to right.
Bottom panel : relaxation time $\tau_\alpha$ for the two systems,
with data from Monte Carlo simulations of a binary mixture
\cite{brambilla} superimposed. The three sets of data are fitted to
the MCT prediction, Eq.(\ref{mct}), using the same exponent
$\gamma=2.7$. The critical volume fraction issued from the fit is
shown in the label.}
\end{figure}

We perform the simulations in two steps. We prepare initial
configurations at the desired state points, and perform a long
equilibration run using periodic velocity rescaling to reach thermal
equilibrium. We then perform a production run in the microcanonical
ensemble. We check very carefully that thermal equilibrium is indeed
reached during equilibration. We also check that our results do not
depend on the duration of these two steps, and we repeat them for at
least five independent samples at each volume fraction to obtain
better statistics. The main issue, mentioned in the original work of
Ref.~\cite{puertas}, is the possibility for the less polydisperse
system ($\sigma \approx 6\%$) to crystallize in the course of the
simulation. In this case the sample is discarded and a new sample
studied. We found that this problem is much more severe than
reported in Ref.~\cite{puertas}, and so we were not able to collect
data above $\phi=0.587$. For the more polydisperse system ($\sigma
\approx 12 \%$), we never detected crystallization and we were able
to collect data up to $\phi=0.60$, where we stopped because we could
not reach equilibrium within our numerical time window. It is
interesting to note that crystallization similarly prevented the
colloidal samples of Ref.~\cite{vanmegen} to be studied at very
large volume fractions, while no such problem is encountered for the
colloidal system studied in this article.

We report in the top panel of Fig.~\ref{tausimu} some representative
results for the self-intermediate scattering function of the system
at two polydispersities, and different volume fractions. This figure
is qualitatively very similar to the experimental data of
Fig.~\ref{figfscaling}, with a pronounced non-exponential relaxation
which slows down dramatically when $\phi$ increases. When volume
fraction is low, we see that the change in polydispersity has no
detectable influence on the dynamics, and data for the two systems
nicely superimpose. A first visible effect of polydispersity is that
we do not have data for the  6~\% polydisperse sample at very large
values of the relaxation time because the system eventually
crystallizes. A second effect can be detected when the glassy regime
is entered. For a fixed $\phi$, the dynamics of the system become
faster when polydispersity is increased, as observed
previously~\cite{poly,sear}.

To quantify these observations, we fit the long time decay of the
relaxation to a stretched exponential form, as in
Eq.~(\ref{EqFitISF}), and show the results for $\tau_\alpha(\phi)$
in the bottom panel of Fig.~\ref{tausimu}. As mentioned before, the
polydispersity has very little effect when dynamics is fast, $\phi <
0.5$, while the more polydisperse sample relaxes faster in the
glassy regime.

We have fitted the dynamic slowing down with increasing $\varphi$ to
the algebraic form suggested by MCT. As expected, we find that
polydispersity affects the value of the critical volume fraction,
although the change is rather modest, since $\phi_c$ increases from
$\phi_c=0.595$ to $\phi_c=0.597$ when changing polydispersity from
6~\% to 12~\%. Note that we used the same exponent $\gamma=2.7$ to
fit both sets of data over a comparable window of relaxation times,
showing that polydispersity does not quantitatively affect the form
of the slowing down. It must be noted that the shift in the absolute
value of $\phi_c$ is much less severe than expected from that for
the random close packing volume fraction on the basis of the
numerical work of Ref.~\cite{refRCPpolydisperse}.

It is interesting to note that deviations from the MCT power law
cannot be detected very clearly for the less polydisperse system,
because crystallization prevents observation of larger relaxation
timescales. For the more polydisperse system, however, we find that
deviations from the MCT power law at even larger volume fractions
become observable, since a broader window of relaxation timescales
can be measured. We believe that similar effects are relevant also
in experiments. As we will show shortly, both the experiments
reported in Ref.~\cite{vanmegen} and our data are described by a
comparable power law over a similar window of relaxation timescales.
However, while previous work did not detect deviations from an
algebraic divergence, by studying a more polydisperse system, we can
efficiently suppress crystallization and access a range of
equilibrium relaxation timescales that are beyond the volume
fraction regime which can be described by MCT, but the MCT regime
itself is not qualitatively affected by polydispersity effects.

For completeness, we report in the bottom panel of
Fig.~\ref{tausimu} the Monte Carlo results obtained in
Ref.~\cite{brambilla} for a three-dimensional 50:50 binary mixture
of hard spheres of diameter ratio 1.4, for which polydispersity is
$\sigma \approx 16.7~\%$. Again, we find that the data is
qualitatively unaffected by the use of a very different particle
size distribution, by the fact that a true hard sphere potential
is used, and by the use of a stochastic microscopic dynamics.
An MCT power law also describes the data over a comparable
time window of about three decades, with a similar exponent
$\gamma=2.7$, and a critical volume fraction $\phi_c = 0.593$.
Remember that we cannot directly compare the result for $\phi_c$ to
the ones for the quasi-hard sphere system, since the latter is
dependent on the temperature scale used (a larger temperature would
yield a smaller critical density). Overall, it is reassuring that
whenever comparison is possible, neither the choice of a microscopic
dynamics, nor the particle size distribution and polydispersity seem
to influence the dynamics in a drastic manner. This certainly
suggests that quantitative comparison between different experimental
samples is meaningful.

\subsection{Comparison with previous work}
\label{SUBSEC:comparison}

Considering colloidal hard spheres from the viewpoint of the
glass transition field, it is very natural to expect,
just as is found for {\it all} molecular glasses, that the dynamic
singularity deduced from fitting data to an algebraic law
predicted by MCT is eventually avoided as the glassy regime
is entered more deeply. Yet, the present experiment
is the first direct demonstration that the MCT dynamic
transition is avoided in a colloidal hard sphere sample.
Although we are tempted to assume that our conclusions should
apply to {\it all} colloidal hard spheres samples, this statement would
seem to require additional discussion.

First, it is reassuring to observe that several computational studies
performed in recent years with hard spheres,
using different types of polydispersities,
and different types of microscopic dynamics indicated the
presence of strong deviations from an algebraic divergence
when approaching $\phi_c$. In that sense, the experiments
of Ref.~\cite{vanmegen} stood as an exception, although arguably
an important one.

In Fig.~\ref{figtauexperiments}, we report data from
Ref.~\cite{vanmegen} ($\tau_{\alpha}^{\mathrm{vM}}$, blue circles
and crosses and right axis), together with our data, taken from
Ref.~\cite{brambilla} ($\tau_{\alpha}$, black squares and left
axis). The open circles are the data as presented in
Ref.~\cite{vanmegen}, using $\phi$ values that may be underestimated
by a factor $\leq 1.04$, because of polydispersity, as discussed in
Sec. \ref{tokuyama}. An MCT fit to
$\tau_{\alpha}^{\mathrm{vM}}(\phi)$ yields $\phi_{\mathrm{c}} =
0.572\equiv \phi_{\mathrm{c}}^{\mathrm{vM}}$~\cite{vanmegen} (blue
line). The crosses are the same data plotted against volume fraction
scaled by a factor $\phi_{\mathrm{c}}/
\phi_{\mathrm{c}}^{\mathrm{vM}}  = 1.031$, with $\phi_{\mathrm{c}}=
0.59$ the critical volume fraction obtained by fitting our data to
the MCT prediction, Eq.~(\ref{mct})~\cite{noteshifttime}. Note that
the scaling factor required for the MCT critical volume fractions to
coincide is smaller than the uncertainty in the absolute $\phi$ for
both sets of data. When compared over the range of volume fraction
where MCT applies, both sets of data are nearly indistinguishable,
confirming that in this regime sample polydispersity has a very
small influence on the dynamics, as indicated by the simulations
discussed in Sec.~\ref{simu}.

Figure~\ref{figtauexperiments} shows also that the range of
timescales reported in Ref.~\cite{vanmegen} is relatively smaller by
about two decades than the present data. Assuming that the behavior
of the fluid at larger $\phi$ for the sample of Ref.~\cite{vanmegen}
would also compare well with the one we observe, one realizes that
the data point of Ref.~\cite{vanmegen} at the highest $\phi$ lies
just below the volume fraction where deviations from MCT predictions
should start becoming visible, possibly explaining why no such
deviations were observed in that work. Of course, obtaining
equilibrium data at larger $\phi$ for this sample would be
difficult, because slow dynamics would start competing with
crystallization, as we observed numerically in Sec.~\ref{simu} for
the sample at the lowest polydispersity.

\section{Conclusions}
\label{SEC:Conclusions}

We have discussed some of the challenges that must be faced when
investigating the slow dynamics of concentrated colloidal systems,
together with the solutions we devised. These challenges include the
determination of the absolute volume fraction and avoiding or at
least minimizing artefacts due to sedimentation and convection. At
very large volume fractions, aging and dynamical heterogeneity make
measurements even more difficult. The data become more erratic,
suggesting that a proper ensemble average is not being achieved.

The data presented here and in~\cite{brambilla} show unambiguously
that, for our sample, the power-law divergence of $\ta(\phi)$ is
avoided and that at high $\phi$ the growth of the relaxation time is
well described by a generalized VFT law. A comparison with the
measurements reported by van Megen \textit{et al.}~\cite{vanmegen}
indicates that, in the MCT regime, the two sets of data are
compatible within the experimental uncertainty on the absolute
volume fraction. This suggests that the slow dynamics of hard
spheres is only weakly sensitive to the precise value of
polydispersity, at least up to moderate values $\sigma~\sim 10\%$.
Our numerical work supports
this conclusion. The agreement between our data and
previous work strongly suggests
that the absence of a true algebraic
divergence of $\ta$ should be a general feature in colloidal hard
spheres, independent of the details of the system. In this respect, hard
spheres should be similar to molecular glass formers, for which the
MCT transition is avoided as well.

Several questions remain open. While we were able to rule out the
existence of an MCT glass transition at $\phi = \phic \approx 0.59$,
our experiments do not allow us to determine unambiguously whether
the location of the divergence of $\ta$ obtained by a generalized
VFT fit, $\phi_0 = 0.637$, is distinct from $\phi_{\mathrm{rcp}}$.
Addressing this issue would allow one to discriminate between
competing theoretical approaches~\cite{jorge}, e.g. recent excluded volume
approaches~\cite{ken} and thermodynamic glass transition
theories~\cite{silvio,zamponi}. However, providing an experimental answer
will be particulary difficult: on one hand, $\phi_0$ can only be
obtained by extrapolation. On the other hand, $\phi_{\mathrm{rcp}}$
is difficult to locate both on a conceptual
standpoint~\cite{torquato2} and because any deviations from the
behavior of an ideal hard sphere potential may become important very
close to $\phi_{\mathrm{rcp}}$.

Another promising line of research concerns the aging regime. We have
shown that at the highest volume fractions investigated the dynamics
slows down over a period of the order of a few (equilibrium)
structural relaxation times. Whether a similar behavior also applies
to more diluted samples remains to be ascertained. Indeed, in our
experiments the thermalization time, during which convective motion
is likely to set in due to small temperature gradients across the
sample, was too long to reach any conclusive result for $\phi <
0.5953$. The large dynamical heterogeneity of the most concentrated
sample suggests that dynamical fluctuations may be much larger than
previously thought in out-of-equilibrium samples at very high
$\phi$. The interplay between dynamical heterogeneity and the
evolution of the dynamics in the aging regime appears as a promising
field for future investigation.

\begin{acknowledgments}
We thank
G. Biroli,
W. Kob,
A. Liu,
P. Pusey,
and V. Trappe
for stimulating discussions and suggestions. We also thank
V. Martinez and W. Poon for raising many issues about Ref.~\cite{brambilla},
which motivated us to reply in some parts of the present report.
This work was supported in
part by the European MCRTN ``Arrested matter''
(MRTN-CT-2003-504712), the European NoE ``Softcomp'', and by CNES
and  the French Minst\`{e}re de la Recherche (ACI JC2076, ANR
``DynHet''). L.C. ackowledges the support of the Institut
Universitaire de France.
\end{acknowledgments}


\begin{thebibliography}{99}

\bibitem{thebook}
W. B. Russel, D. A. Saville, and W. R. Schowalter,
{\it Colloidal dispersions} (Cambridge University Press, Cambridge, 1992).

\bibitem{HabdasCOCIS2002} P. Habdas and E. R. Weeks, Current Opinion in Colloid \& Interface Science \textbf{7}, 196 (2002).

\bibitem{Berne1976} B. J. Berne and R. Pecora, \textit{Dynamic Light Scattering} (Wiley, New York, 1976).

\bibitem{PuseyNature1986} P. N. Pusey and W. van Megen, Nature \textbf{320}, 340 (1986).

\bibitem{PoonJPCM2002} W. C. K. Poon, J. Phys.: Condens. Matter \textbf{14}, R859 (2002).

\bibitem{GlotzerNatMat2007} S. C. Glotzer and M. J. Solomon, Nature Materials \textbf{6}, 557 (2007).

\bibitem{KegelLangmuir2000} W. K. Kegel, Langmuir \textbf{16}, 939 (2000).

\bibitem{SimeonovaPRL2004} N. B. Simeonova and W. K. Kegel, Phys. Rev. Lett. \textbf{93}, 035701 (2004).


\bibitem{brambilla}
G. Brambilla, D. El Masri, M. Pierno, G. Petekidis, A. B. Schofield,
L. Berthier, and L. Cipelletti, Phys. Rev. Lett. {\bf 102}, 085703 (2009).


\bibitem{mct} W. G\"otze, J.
Phys.: Condens. Matter {\bf 11}, A1 (1999).

\bibitem{vanmegen} W. van Megen, T. C. Mortensen, S. R. Williams, J. M\"uller, Phys. Rev. E {\bf 58}, 6073 (1998).

\bibitem{pusey2}  P. N. Pusey, and W. Van Megen, Phys. Rev. Lett. {\bf 59}, 2083 (1987).

\bibitem{weitz}
T. G. Mason and D. A. Weitz,
Phys. Rev. Lett. {\bf 75}, 2770 (1995).

\bibitem{pusey} P. N. Pusey, and W. Van Megen,
Nature {\bf 320}, 595 (1986).

\bibitem{free} M. H. Cohen and D. Turnbull,
J. Chem. Phys. {\bf 31}, 1164 (1959).

\bibitem{ken} K. S. Schweizer,
J. Chem. Phys. {\bf 127}, 164506 (2007).

\bibitem{silvio} M. Cardenas, S. Franz, and G. Parisi,
J. Chem. Phys. \textbf{110}, 1726
(1999).

\bibitem{zamponi}
G. Parisi and F. Zamponi, J. Chem. Phys. {\bf 123}, 144501 (2005).

\bibitem{chaikin} Z. Cheng, J. Zhu, P. M. Chaikin, S. Phan, and W. B. Russel, Phys. Rev. E {\bf 65}, 041405 (2002).

\bibitem{szamel} S. K. Kumar, G. Szamel, and J. F. Douglas,
J. Chem. Phys. {\bf 124}, 214501 (2006).

\bibitem{reviewnature}
P. G. Debenedetti and F. H. Stillinger, Nature {\bf 410}, 259
(2001).

\bibitem{pointJ}
C. S. O'Hern, S. A. Langer, A. J. Liu, and S. R. Nagel, Phys. Rev.
Lett. {\bf 88} 075507 (2002).

\bibitem{torquato2} S. Torquato, T. M. Truskett, and P. G.
Debenedetti, Phys. Rev. Lett. {\bf 84}, 2064 (2000).

\bibitem{longtom} L. Berthier and T. A. Witten, arXiv:0903.1934.

\bibitem{wolynes} T. R. Kirkpatrick, D. Thirumalai, and P. G. Wolynes,
Phys. Rev. A {\bf 40}, 1045 (1989).

\bibitem{BB} J.-P. Bouchaud and G. Biroli,
J. Chem. Phys. {\bf 121}, 7347 (2004).

\bibitem{Chaikin2} S.-E. Phan, W. B. Russel, Z. Cheng, J. Zhu,
P. M. Chaikin, J. H. Dunsmuir, R. H. Ottewill, Phys. Rev. E
{\bf 54}, 6633 (1996).

\bibitem{brownian}  S. P. Das and G. F. Mazenko,
Phys. Rev. A {\bf 34}, 2265 (1986).

\bibitem{ABL1} M. E. Cates and S. Ramaswamy,
Phys. Rev. Lett. \textbf{96}, 135701 (2006).

\bibitem{ABL2}
A. Andreanov, G. Biroli, and A. Lef\`evre, J. Stat. Mech. P07008
(2006).

\bibitem{stochastic1} G. Szamel and E. Flenner, Europhys. Lett. {\bf 67},
779 (2004).

\bibitem{stochastic2} L. Berthier and W. Kob,
J. Phys.: Condens. Matter {\bf 19}, 205130 (2007).

\bibitem{stochastic3}
L. Berthier,
Phys. Rev. E {\bf 76}, 011507 (2007).



\bibitem{PuseyJPhysChem1982} P. N. Pusey, J. Phys. A 11, 119 (1978);
P. N. Pusey, H. M. Fijnaut, and A. Vrij, J. Chem. Phys.
\textbf{77}, 4270 (1982).

\bibitem{refRCPpolydisperse}
W. Schaertl and H. Sillescu, J. Stat. Phys. {\bf 74},
1007 (1994).

\bibitem{ElMasriJPCM2005} D. El Masri, M. Pierno, L. Berthier, and L. Cipelletti, J. Phys.: Condens. Matter \textbf{17}, S3543 (2005).

\bibitem{LucaJPCM2003} L. Cipelletti, H. Bissig, V. Trappe, P. Ballesta, and S. Mazoyer, J. Phys.: Condens. Matter \textbf{15}, S257 (2003).

\bibitem{DuriPRE2005} A. Duri, H. Bissig, V. Trappe, and L. Cipelletti, Phys. Rev. E \textbf{72}, 051401 (2005).

\bibitem{XuePRA1992} J. Z. Xue, D. J. Pine, S. T. Milner, X. L. Wu, and P. M. Chaikin, Phys. Rev. A \textbf{46}, 6550 (1992).

\bibitem{Goodman1975} J. W. Goodman, in \textit{Laser speckles and related phenomena}, edited by J. C. Dainty (Springer-Verlag, Berlin, 1975), Vol. 9, p. 9.

\bibitem{Tokuyama} M. Tokuyama, and I. Oppenheim, Phys. Rev. E {\bf 50}, R16 (1994).

\bibitem{Beenakker} C. W. J. Beenakker and P.Mazur, Physica A {\bf 120} 388 (1983).

\bibitem{SegrePRE1995} P. N. Segre, O. P. Behrend, and P. N. Pusey, Phys. Rev. E \textbf{52}, 5070 (1995).

\bibitem{bellon}
L. Bellon, M. Gibert, and R. Hernandez,
Eur. Phys. J. B {\bf 55}, 101 (2007).

\bibitem{CoussotPRL2002} P. Coussot, Q. D. Nguyen, H. T. Huynh, and D. Bonn, Phys. Rev. Lett. \textbf{88}, (2002).

\bibitem{sgr} P. Sollich, F. Lequeux, P. Hebraud, and M. E. Cates,
Phys. Rev. Lett. {\bf 78}, 2020 (1997).

\bibitem{2time} L. Berthier, J.-L. Barrat, and J. Kurchan,
Phys. Rev. E {\bf 61}, 5464 (2000).

\bibitem{TokumaruExpInFluids1995} P. T. Tokumaru and P. E. Dimotakis, Experiments in Fluids \textbf{19}, 1 (1995).

\bibitem{NobachExpFluids2005} H. Nobach, N. Damaschke, and C. Tropea, Experiments in Fluids \textbf{39}, 299 (2005).

\bibitem{NicolaiPhysFluids1995} H. Nicolai, B. Herzhaft, E. J. Hinch, L. Oger, and E. Guazzelli, Physics of Fluids \textbf{7}, 12 (1995).

\bibitem{CipellettiJPCM2005} L. Cipelletti and L. Ramos, J. Phys.: Condens. Matter \textbf{17}, R253 (2005).


\bibitem{KirschJChemPhys1996} S. Kirsch, V. Frenz, W. Schartl, E. Bartsch, and H. Sillescu, J. Chem. Phys. \textbf{104}, 1758 (1996).

\bibitem{LucaRSI1999} L. Cipelletti and D. A. Weitz, Rev. Sci. Instrum. \textbf{70}, 3214 (1999).

\bibitem{ViasnoffRSI2002} V. Viasnoff, F. Lequeux, and D. J. Pine, Rev. Sci. Instrum. \textbf{73}, 2336 (2002).

\bibitem{BerthierScience2005} L. Berthier, G. Biroli, J. P. Bouchaud, L. Cipelletti, D. El Masri, D. L'Hote, F. Ladieu, and M. Pierno, Science \textbf{310}, 1797 (2005).

\bibitem{dalleferrier} C. Dalle-Ferrier,
C. Thibierge, C. Alba-Simionesco, L. Berthier, G. Biroli, J.-P. Bouchaud,
F. Ladieu, D. L'H\^ote, and G. Tarjus,
Phys. Rev. E {\bf 76}, 041510 (2007).

\bibitem{WeeksScience2000} E. R. Weeks, J. C. Crocker, A. C. Levitt, A. Schofield, and D. A. Weitz, Science \textbf{287}, 627 (2000).

\bibitem{WeeksJPCM2007} E. R. Weeks, J. C. Crocker, and D. A. Weitz, J. Phys.: Condens. Matter \textbf{19}, (2007).

\bibitem{EdigerReview} M. D. Ediger, Annu. Rev. Phys. Chem. \textbf{51}, 99 (2000).

\bibitem{GlotzerReview} S. C. Glotzer, J. Non-Cryst. Solids \textbf{274}, 342 (2000).

\bibitem{foampeter}
P. Mayer, H. Bissig, L. Berthier, L. Cipelletti, J.P. Garrahan,
P. Sollich, and V. Trappe,
Phys. Rev. Lett. {\bf 93}, 115701 (2004).

\bibitem{BallestaNatPhys2008} P. Ballesta, A. Duri, and L. Cipelletti, Nature Physics \textbf{4}, 550 (2008).

\bibitem{tom} L. Berthier and T. A. Witten, arXiv:0810.4405.

\bibitem{poly}
S. R. Williams and W. van Megen,
Phys. Rev. E {\bf 64}, 041502 (2001).

\bibitem{DjamelPhD} D. El Masri, PhD thesis, Universit\'{e}
Montpellier 2 (2007).

\bibitem{puertas} T. Voigtmann, A. M. Puertas, and M. Fuchs,
Phys. Rev. E {\bf 70}, 061506 (2004).

\bibitem{allen} M. Allen and D. Tildesley,
{\it Computer Simulation of Liquids} (Oxford University Press,
Oxford, 1987).

\bibitem{sear} R. P. Sear, J. Chem. Phys.
{\bf 113}, 4732 (2000).

\bibitem{jorge} F. Krzakala and J. Kurchan,
Phys. Rev. E {\bf 76}, 021122 (2007).

\bibitem{noteshifttime} A shift factor on the $y$ axis was used to
match the datasets. This factor accounts for the differences in
particle size, solvent viscosity, $q$ vector and definition of $\ta$
in the two experiments.

\end{thebibliography}
\end{document}